\documentstyle[amsfonts,preprint,aps]{revtex}

\begin{document}
\draft
\preprint{ }

\title{Critical behavior of a fluid in a disordered porous matrix.\\
An Ornstein-Zernike approach.}
\author {E. Pitard$^1$, M.L. Rosinberg$^1$, G. Stell$^2$ and G. Tarjus$^1$}
\address{
$^1$Laboratoire de Physique Th\'eorique des Liquides \cite{AAAuth},
Universit\'e Pierre et Marie Curie,\\
4 Place Jussieu, 75252 Paris Cedex 05, France}
\address{$^2$Department of Chemistry, State University of New York at
 Stony Brook, \\
Stony Brook, NY 11794, USA}
 \date{\today}
\maketitle
\begin{abstract}
Using a liquid-state approach based on Ornstein-Zernike equations, we study the
behavior of a fluid inside a porous disordered matrix
near the liquid-gas critical point. The results obtained within various
standard approximation schemes such as lowest-order $\gamma$-ordering and the
mean-spherical approximation suggest that the critical behavior is closely
related to that of the random-field Ising model (RFIM).
\end{abstract}
\pacs{Pacs numbers: 05.70.Fh, 61.20.Gy, 64.60.Fr}
\narrowtext
\newpage

The interpretation of experimental studies of the phase behavior of fluids
(binary mixtures \cite{MGPL1984} or $^4$He \cite{WC1990})
 adsorbed in disordered porous materials such as silica gels or Vycor remains
highly controversial. Some results
lend support to the view that preferential adsorption of one component and
randomness of the pore network give rise to a random-field Ising-like behavior
near
the critical point, as suggested by Brochard and de Gennes \cite {BdeG1983}. On
the other hand, it has been argued that the experiments in Vycor,
which is a glass with a fairly low porosity, can be interpreted in terms of
wetting in a confined geometry, with no
randomness \cite{MLG1992}. Clear-cut conclusions are somewhat obscured by the
metastable and non-equilibrium effects which are often present in the
experiments.
However, there is a need for more realistic continuum descriptions which
encompass both randomness, confinement,
and connectivity between pores.  The goal of this letter is to present a
liquid-state approach describing a single
component fluid in a rigid disordered matrix and, starting from this
microscopic theory, to give some insight
on the critical properties of the fluid and their relation to the random-field
Ising model (RFIM) \cite{NV1988}.

Following the formalism recently proposed by Madden and Glandt \cite{MG1988},
we consider the matrix-fluid system
as a special binary mixture in which the matrix is treated as a rigid set of
obstacles, obtained by quenching an equilibrium
configuration of particles (species 0), with the fluid particles (species 1) at
equilibrium in the presence of (and in interaction
with) the quenched particles. The structure of this quenched-annealed (Q-A)
mixture is described by the total pair
correlation functions $h_{ij}(r)$ which satisfy a set of coupled integral
equations, the so-called replica Ornstein-Zernike
 (ROZ) equations \cite{GS1992}. As in the usual OZ formalism \cite {HMcD1976},
these equations relate the functions $h_{ij}(r)$
to the corresponding direct correlation functions $c_{ij}(r)$ which are
expected to be short-ranged (i.e, $\int d^d{\bf r} \,c_{ij}(r)$ is finite),
even at the critical point.
The exact form of the ROZ equations can be most easily derived by using the
replica trick which replaces the study of the original  system by that of an
equilibrium mixture composed of $s$ copies of the fluid interacting with the
matrix. It should be stressed that, in contrast to
the standard replica treatment of quenched disorder, there is no direct
interaction between the fluid replicas because the average over disorder (i.e.,
over all possible configurations of the matrix particles) is not performed
explicitly.
These  ROZ equations represent the starting point of our treatment and their
formal solution in Fourier space is given by

\begin{mathletters}
\begin{equation}
\hat{h}_{00}(k)=\frac{\hat{c}_{00}(k)}{1-\rho_0 \hat{c}_{00}(k)}
\end{equation}

\begin{equation}
\hat{h}_{01}(k)=\frac{\hat{c}_{01}(k)}{[1-\rho_0 \hat{c}_{00}(k)][1-\rho_1
\hat{c}_c(k)]}
\end{equation}

\begin{equation}
\hat{h}_c(k)=\frac{\hat{c}_c (k)}{1-\rho_1 \hat{c}_c(k)}
\end{equation}

\begin{equation}
\hat{h}_b(k)=[\hat{c}_b(k)+\rho_0\frac{\hat{c}_{01}^2(k)}{1-\rho_0
\hat{c}_{00}(k)}]\frac{1}{[1-\rho_1 \hat{c}_c(k)]^2}
\end{equation}
\end{mathletters}
where $\hat{f}({\bf k})$ denotes the Fourier transform of a function $f({\bf
r})$ and $\rho_i$ is the average density of species
$i$. In deriving these equations, it has been assumed that there is no replica
symmetry breaking. As in the case of
disordered magnetic systems, one has introduced the connected part, $h_c(r)$,
and the blocked or
 disconnected part, $h_b(r)=h_{11}(r)-h_c(r)$, of the fluid-fluid correlation
function \cite {GS1992,RTS1994}; they are equal, respectively, to
 the $s=0$ limits of $h_{\alpha \alpha}(r) -h_{\alpha \beta}(r)$ and
$h_{\alpha \beta}(r)$, where $\alpha$ and $\beta \neq \alpha$ denote fluid
replicas.  Similarly, $c_c(r)$ and  $c_b(r)$ are defined as the connected and
blocked parts of $c_{11}(r)$. By symmetry, one has $h_{01}=h_{10}$ and
$c_{01}=c_{10}$.

The above ROZ equations differ from those recently derived for a fluid in
presence of a quenched,
random, external potential obeying a Gaussian distribution \cite {MD1994}:  in
Eq. (1d), the quenched disorder enters explicitly
 through a term involving the matrix-matrix and  matrix-fluid direct
correlation functions which also appear in Eqs. (1a) and (1b), whereas
 in Ref. \cite {MD1994} it is only contained in the inter-replica pair
potential. As a result, the present formulation is also able to take into
account the geometrical constraint on the fluid due to the presence of the
matrix (excluded volume and/or confinement effects).

It has been shown \cite{RTS1994,FG1994} that $\hat{c}_c(0)$ is related to
$\chi_1$, the compressibility of the fluid inside the matrix (i.e., the
derivative of the fluid pressure $P_1$ with respect to the density or, through
the Gibbs-Duhem equation \cite {RTS1994}, the derivative of the fluid chemical
potential), via the equation

\begin{equation}
k_BT\rho_1 \chi_1=[1-\rho_1 \hat{c}_c(0)]^{-1}
\end{equation}
Therefore, the gas-liquid critical point of the fluid inside the matrix, if it
exists, is characterized as usual by both
a diverging compressibility and long-range correlations. However, a special
feature of Q-A systems
 is that the disconnected part of the fluid-fuid correlation function, and as a
consequence $h_{11}(r)$
itself, may be much longer-ranged that the connected part. Indeed, if one
defines
the exponents $\eta$ and $\overline{\eta}$ by $h_c(r) \sim r^{-d+2-\eta}$ and
$h_b(r) \sim r^{-d+4-
\overline{\eta}}$ when $r \rightarrow \infty$ at the critical point, one finds
immediately from Eqs. (1c) and (1d) that $\overline{\eta}=2\eta$.
This result rests only on the assumption that the matrix is non-critical, i.e.,
$1-\rho_0 \hat{c}_{00}(0) \neq 0$,  and that $\hat{c}_b(0)$ and
 $\hat{c}_{01}(0)$ are finite, a condition generally expected for direct
correlation functions. This relation between $\overline{\eta}$ and
$\eta$ is for instance satisfied within the standard Ornstein-Zernike
approximation which assumes that the direct correlation functions have the same
range as the associated pair interactions. For short-ranged interactions, this
gives $\overline{\eta}=\eta=0$, but for long-ranged fluid-fluid interactions
decaying as $r^{-(d+\sigma)}$
for $r \rightarrow \infty $ with $\sigma\leq 2$, one obtains
$\overline{\eta}=4-2\sigma=2\eta$. Note that the same relation
$\overline{\eta}=2\eta$ has been proposed for the RFIM  \cite{S1985} and seems
to be confirmed by recent numerical estimates \cite{GAAHS1993}.

To be more specific, we now suppose that the fluid-fluid (resp. matrix-fluid)
pair potential can be divided into a reference part
which includes the short-range repulsive part of the interaction and a more
smoothly varying long-range attractive part  $w_{11}(r)$ (resp.
$w_{01}(r)$).
As a first step, we derive the mean-field (MF) results. As in the bulk fluid
case, this can be done by considering  the lowest $\gamma$-
ordering approximation which is derived by introducing the inverse range
parameter $\gamma$ in the attractive fluid-fluid potential,
 $w_{11}(r)=\gamma^d \phi(\gamma r)$, and  then taking the limit $\gamma
\rightarrow 0$ (see for instance Ref. \cite{HMcD1976}) .
{}From the graphical expansion of the three direct correlation functions
$c_{01}(r)$, $c_b(r)$ and $c_c(r)$ \cite{MG1988}, it is easy to show
that $\hat{c}_{01}(k) =\hat{c}_{01}^R(k)$, $\hat{c}_b(k) =\hat{c}_b^R(k)$ and
$\hat{c}_c(k) =\hat{c}_c^R(k)- (k_BT)^{-1}\hat{\phi}(0) \delta_{k,0}$, where
$\delta_{k,0}$ is the Kronecker symbol and the superscript $R$ indicates a
quantity calculated in a reference
system for which $w_{11}(r) \equiv 0$. The fact that $\hat{c}_b$ reduces to its
reference part illustrates the fact that the fluid replicas are not
directly coupled (see above). Using Eq. (2), we immediately obtain a van der
Waals equation of state,
 $P_1=P_1^{(R)} + \frac{1}{2} \rho_1^2 \hat{\phi}(0)$.
{}From the thermodynamic relations satisfied by the Q-A mixture \cite{RTS1994},
we can conclude that besides $\eta=\overline{\eta}=0$, the critical behavior is
described by the usual MF
exponents, $\nu=1/2$, $\alpha=0$, $\beta=1/2$, $\gamma=1$, $\delta=3$. As
illustrated numerically in Ref. \cite{FG1994b}
for a specific system,
the critical point is displaced from the MF critical point of the bulk fluid
because the reference system
  includes the influence of the host matrix
(note that in Ref. \cite{FG1994b} the matrix-fluid interaction is also treated
in the MF approximation).

To go further, some more quantitative statement on the correlations over small
distances must be made. This can be implemented, within the
OZ approximation mentioned above, by considering approximate closure relations
to the ROZ equations like the
mean-spherical approximation (MSA) \cite {HMcD1976}. For bulk fluids, the MSA
is known to
yield the same critical exponents as the spherical spin model \cite {S1969}.
We now show that the same is true for a Q-A system.
To avoid unnecessary technical difficulties, we consider the case of a
d-dimensional lattice gas.
We expect that the behavior in the critical region will be identical to that of
a real
continuum fluid. Lattice cells (or sites) can be occupied either by matrix or
fluid "particles" and the pair potentials contain a " hard-core"
 part which excludes multiple occupancy of a cell (or site).  It must be noted
that even for the lattice
 fluid there is no hole-particle symmetry because of
the interaction with the matrix. Therefore, there is no exact correspondence
between this microscopic description of a fluid
 inside a matrix and the usual RFIM (see also Ref. \cite{SP1992} for a related
discussion). More precisely, the standard equivalence
between the lattice-gas and the spin $1/2$ model shows that we are actually
dealing with a site diluted Ising model where the spins are
coupled to correlated external random fields which depend on the site disorder
variables \cite{PRST1994}.
Application of the MSA closure in the matrix-replicas mixture  (i.e. $k_BT
c_{\alpha \beta}({\bf r}) =-w_{\alpha \beta}({\bf r})$ for ${\bf r} \neq 0$ )
gives,
 in the limit $s \rightarrow 0$,

\begin{eqnarray}
 c_{00}({\bf r}) &=&- \frac{\delta_{{\bf r},0}}{(1-\rho_0)},\quad  c_{01}({\bf
r})=c_{01}^0\delta_{{\bf r},0}-\beta w_{01}({\bf r})\nonumber\\
c_b({\bf r})&=&0, \quad c_c({\bf r})=c_{c}^0\delta_{{\bf r},0}-\beta
w_{11}({\bf r}),
\end{eqnarray}
where $c_{01}^0$ and  $c_{c}^0$ are functions of temperature and fluid density,
which, owing to the structure of the ROZ equations,
are completely determined by imposing the "core"  conditions $h_{01}({\bf
r}=0)=h_{11}({\bf r}=0)=-1$. Note that $h_c$
and $h_b$ do not satisfy themselves the core requirement. On the other hand,
the prescription for $c_{00}$ is the direct
consequence of the core condition for the matrix particles, $h_{00}({\bf
r}=0)=-1$ (for simplicity, we assume that there is no attractive
 interaction between matrix particles).

For clarity, we first consider nearest-neighbor (n.n.) interactions in a
hypercubic lattice. Generalizing the analysis
developped by Stell \cite {S1969} and Theumann \cite {T1970} for the bulk
lattice-gas, we write the inverse Fourier transform of
$\hat{\chi}({\bf k})=1+\rho_1 \hat{h}_c({\bf k})$ as

\begin{eqnarray}
-\frac{\rho_1 w_{11}\chi({\bf r})}{k_BT}\equiv G({\bf
r},\kappa^2)=\int_{-\pi}^{\pi} \frac{d^d{\bf k}}{(2\pi)^d}\frac{e^{i {\bf
k}.{\bf r}}}{\kappa^2+2\sum_{j=1}^d (1-\cos k_j)}
\end{eqnarray}
where ${\bf k}=(k_1,...,k_d)$ and $\kappa=[-2d +k_BT(1-\rho_1 c_c^0)/(- \rho_1
w_{11})]^{1/2}$ is an inverse correlation length which vanishes at the critical
point. By replacing in the ROZ equations, we find

\begin{eqnarray}
\frac{-\rho_1 w_{11}}{(1-\rho_0)}h_{01}({\bf r})=fG({\bf
r},\kappa^2)+w_{01}\delta_{{\bf r},0}
\end{eqnarray}
and

\begin{eqnarray}
\frac{(\rho_1 w_{11})^2}{\rho_0(1-\rho_0)}h_b({\bf r})=w_{01}^2+2w_{01}fG({\bf
r},\kappa^2)-f^2 G'({\bf r},\kappa^2)
\end{eqnarray}
where $G'({\bf r},\kappa^2)\equiv \partial G({\bf r},\kappa^2)/\partial
\kappa^2$ and $f=k_BTc_{01}^0-w_{01}(\kappa^2+2d)$.
Obviously, $h_c({\bf r})$, $h_b({\bf r})$ and $h_{01}({\bf r})$ have the same
correlation length
$\kappa^{-1}$, and $\eta=\overline{\eta}=0$. If $d \leq 4$, no finite critical
temperature exists since this
would lead to the unacceptable result that $h_b({\bf r})$ and thus $h_{11}({\bf
r})$ do not decay to zero when $r \rightarrow  \infty$ at the
critical point. Using the core conditions together with Eqs. (4-6) yields an
implicit equation relating $\kappa$ to the thermodynamic variables. In absence
of hole-particle symmetry, the critical point is found by requiring that
$(\partial P_1/\partial \rho_1)_c=(\partial^2 P_1/\partial \rho_1^2)_c=0$ on
the critical isotherm,  which from Eqs. (2) and (4) is equivalent to
$\kappa_c=(\partial \kappa^2/\partial \rho_1)_c=0$. For our system, this yields
$\rho_{1c}=(1-\rho_0)(1+2yq\rho_0)/2(1+q\rho_0)$ with $q=-1-G'(0,0)/G^2(0,0)
\geq 0$ and $y=w_{01}/w_{11}$, whereas the shift of the critical temperature
due to the presence of the matrix
is given by $T_c/T_c^{bulk}=(1-\rho_0) [1+4y(1-y)q\rho_0]/[1+q\rho_0]$; this
ratio being always smaller than $1$, one has $T_c \leq T_c^{bulk}$.
The behavior of the correlation length in the critical regime can now be
obtained by expanding the implicit equation for $\kappa$
 in terms of $\Delta T=T- T_c$, $\Delta \rho_1=\rho_1-\rho_{1c}$, and $\kappa$.
Using the
 asymptotic expansion of the Green function $G({\bf r}=0,\kappa^2)$ \cite
{S1969}, we finally obtain for $d > 4$

\begin{eqnarray}
-\frac{G(0,0)}{w_{11}}k_B\Delta T &+&\frac{1+q\rho_0}{1-\rho_0}(\Delta
\rho_1)^2=k_BT_c \frac{G'(0,0)}{w_{11}}\kappa^2\nonumber\\
&+&\frac{\rho_0(1-\rho_0)}{4(1+q\rho_0)^2}(1-2y)^2[A\kappa^2+B\kappa^{d-4}+C\kappa^{d-4}\ln \kappa+...]
\end{eqnarray}
where A, B are nonzero constants and $C=0$ when and only when d is odd. From
this we can read off the exponents of the correlation
 length
along the critical isochore ($\nu $) and the critical isotherm ($\epsilon $, in
the notation of Ref. \cite{S1969}), and then, with the help
of the compressibility equation (2) and the Gibbs-Duhem relation
\cite{RTS1994}, the values of $\gamma $ and $\delta $.
For $d \geq 6$, one recovers the MF exponents whereas for $4< d \leq 6$ one has
$\nu=1/(d-4)$, $\epsilon=2/(d-4)$,
 $\gamma=2/(d-4)$ and $\delta=d/(d-4)$. On the other hand, when $y=1/2$, one
finds the MF exponents as
soon as $d \geq 4$. In this latter case, the system is easily shown to be
isomorphic to the site diluted spin model with no random fields.

Using the expression of the configurational internal energy, $U_1= 1/2
\rho_1\hat{w}_{11}(0)+\rho_0\hat{w}_{10}(0)
+1/2\sum_{{\bf r}\neq 0} [\rho_1 w_{11}({\bf r})h_{11}({\bf r})+2\rho_0
w_{10}({\bf r})h_{10}({\bf r})]$,
and deriving with respect to temperature, one finds that the critical exponent
$\alpha$ of the fluid specific heat  $C_{1v}$
is always zero whereas the exponent $\alpha_s$ characterizing the singular part
of $C_{1v}$, if present,  is always negative
and given by $\alpha_s=(d-6)/(d-4)$ for $4<d<6$ and $\alpha_s=(6-d)/2$ for $d
\geq 6$.
Finally, the same arguments as for the bulk lattice-gas \cite {S1969} lead to
$\beta=1/2$ for the coexistence curve exponent.
These calculations can be easily generalized to
 non n.n. interactions and to cases where  $w_{01}(r)$ and  $w_{11}(r)$ do not
have the same range. The same exponents are recovered
except  that when $w_{11}(r)$ dies off like $r^{-(d+\sigma)}$ with $\sigma< 2$
one finds $\nu=1/(d-2\sigma)$, $\gamma=\sigma/(d-2\sigma)$,
$\delta=d/(d-2\sigma)$ and $\alpha_s=(d-3\sigma)/(d-2\sigma)$
for $2 \sigma< d < 3 \sigma $, whereas $\nu=1/\sigma$, $\gamma=1$, $\delta=3$
and $\alpha_s=(3\sigma-d)/\sigma$ for $d \geq 3\sigma$.

Except for the special case which reduces to the site diluted spin model
without random field,
all  these exponents coincide with those of the spherical model in a random
field \cite{LT1974}: in
particular, we find that the disorder induces a dimensional shift $d
\rightarrow d-2$ for short-range interactions and $ d \rightarrow d-\sigma$ for
long-range interactions. These results are valid even in the absence of
attractive matrix-fluid interaction ($y=0$ in the case of n.n interactions).
The fact  that the Q-A system and the random-field spin model have the same
behavior in the critical region does not garantee,
however, that their thermodynamics coincide, and a more complete study of the
MSA phase diagram will be presented elsewhere
 \cite {PRST1994}.

To summarize, we have related the critical behavior of a fluid in a porous
matrix to that of the RFIM by using
common approximations of liquid-state theory  \cite{comment}. The present
treatment based on the ROZ equations takes into account both
disorder and excluded volume effects. The fact that the connection with the
RFIM holds also in the absence
 of matrix-fluid attractive interactions
suggests that preferential adsorption may not be a crucial ingredient for
observing a random-field like behavior.
Of course, the identity of the critical exponents within various approximation
schemes is not a proof that the two systems belong to the same universality
class \cite {comment1}. To go further, one must implement
 the techniques of the renormalization group, which can be considered in
connection with liquid-state theory \cite {PMR1989}.
One must also investigate a possible replica symmetry breaking (RSB) mechanism
which seems to occur in the RFIM
 \cite {MY1992} (however, this is not expected at the MF \cite {SP1977} or
spherical-model levels \cite {MY1992}).  For that purpose,
 it is necessary to use a non-linear closure to the ROZ equations, such as the
hypervertex  approximations discussed in Ref. \cite{SLBT1966}.
If RSB indeed occurs in Q-A systems, the assumption that the direct correlation
functions remain short-ranged,
even in the critical region, may lead to non trivial predictions.

$\quad$

G. S. gratefully acknowledges the support of the National Science Foundation.


\begin{references}
\bibitem[*]{AAAuth} Unit\'e de Recherche Associ\'ee au CNRS (URA 765).
\bibitem{MGPL1984} M.C. Goh, W. I. Goldburg, and C. Knobler, Phys. Rev. Lett.
{\bf 58}, 1008 (1987); S. B. Dierker and P. Wiltzius, {\it ibid.} {\bf 58},
1865 (1987);  S. B. Dierker and P. Wiltzius, {\it ibid.} {\bf 66}, 1185 (1991);
B. J. Frisken and D. S. Canell, {\it ibid.} {\bf 69}, 632 (1992).
\bibitem{WC1990} A. P. Y. Wong and M. H. W. Chan, Phys. Rev. Lett.  {\bf 65},
2567 (1990).
\bibitem{BdeG1983} F. Brochard and P. G. de Gennes, J. Phys. Lett. (Paris)
{\bf 44}, 785 (1983); P. G. de Gennes, J. Phys. Chem. {\bf 88}, 6469 (1984).
\bibitem{MLG1992} L. Monette, A. Liu and G. S. Grest, Phys. Rev. A {\bf 46},
7664 (1992).
\bibitem{NV1988} for recent reviews see T. Nattermann and J. Villain, Phase
Transitions, {\bf 11}, 817 (1988); D. P. Belanger and A. P. Young,
 J. Magn. \&\ Magn. Mater. {\bf 100}, 272 (1991).
\bibitem{MG1988} W. G. Madden and E. D. Glandt, J. Stat. Phys. {\bf 51}, 537
(1988).
\bibitem{GS1992} J. A. Given and G. Stell, J. Chem. Phys. {\bf 97}, 4573
(1992); E. Lomba, J. A. Given, G. Stell, J.J. Weis, and D. Levesque,
Phys. Rev. E {\bf 48}, 223 (1993).
\bibitem{HMcD1976} J. P. Hansen and I. R. McDonald, {\it Theory of Simple
Liquids} (Academic, New York, 1976).
\bibitem{RTS1994} M. L. Rosinberg, G. Tarjus, and G. Stell, J. Chem. Phys. {\bf
100}, 5172 (1994).
\bibitem{MD1994} G. I. Menon and C. Dagupta, Phys. Rev. Lett. {\bf 73}, 1023
(1994).
\bibitem{FG1994} D. M. Ford and E. Glandt , J. Chem. Phys. {\bf 100}, 2391
(1994).
\bibitem{S1985} M. Schwartz, J. Phys. C {\bf 18}, 135 (1985).
\bibitem{GAAHS1993} M. Gofman, J. Adler, A. Aharony, A. B. Harris, and M.
Schwartz, Phys. Rev. Lett.  {\bf 71}, 1569 (1993).
\bibitem{FG1994b} D. M. Ford and E. Glandt , Phys. Rev. E {\bf 50}, 1280
(1994).
\bibitem{S1969} G. Stell, Phys. Rev. {\bf 184}, 135 (1969); G. Stell, in {\it
Phase transitions and critical phenomena},
edited by C. Domb and M. S. Green (Academic, London, 1976), Vol. 5b, p. 205.
\bibitem{SP1992} D. Stauffer and R. B. Pandey, J. Phys. A {\bf 25}, L1079
(1992).
\bibitem{PRST1994} E. Pitard, M. L. Rosinberg, G. Stell, and G. Tarjus, in
preparation.
\bibitem{T1970} W. K. Theumann, Phys. Rev. B {\bf 2}, 1396 (1970).
\bibitem{LT1974} P. Lacour-Gayet and G. Toulouse, J. Phys. (Paris), {\bf 35},
425 (1974).
\bibitem{comment} We have also calculated the critical behavior of the Q-A
mixture within  the generalized mean spherical approximation (GMSA)  in its
self-consistent version  (J. S. Hoye and G. Stell, Mol. Phys. {\bf 52}, 1071
(1984)): one then recovers the exponents of the Gaussian
model of spins in a random field \cite {LT1974}.
\bibitem{comment1} However, since the usual thermodynamic relations remain
valid in a Q-A system \cite{RTS1994}, one can repeat Widom's scaling approach
of the thermodynamic potentials in terms of the reduced fluid variables to get
$\alpha+2\beta+\gamma=2$, $\gamma=\beta (\delta-1)$, as well as
$\gamma=(2-\eta)\nu$. Moreover, a naive scaling ansatz for the disconnected
part of the fluid-fluid pair density in the two-phase region yields the
relation $2 \beta=(d-4-\overline{\eta})\nu$. Therefore, one has
$2-\alpha=[d-(2-\overline{\eta}+\eta)]\nu$ and the usual hyperscaling relation
is modified, the dimension of the system being apparently shifted by
$2-\overline{\eta}+\eta$ (or $2-\eta$, if we use the relation
$\overline{\eta}=2\eta$). All these relations between exponents, including the
modified hyperscaling relation are the same as in the RFIM \cite {NV1988}.
\bibitem{PMR1989} See e.g., A. Parola, A. Meroni and L. Reatto, Phys. Rev.
Lett. {\bf 62}, 2981 (1989).
\bibitem{MY1992} M. Mezard and A. P. Young, Europhys. Lett. {\bf 18}, 653
(1992).
\bibitem{SP1977} T. Schneider and E. Pytte, Phys. Rev. B {\bf 15}, 1519 (1977).
\bibitem{SLBT1966} G. Stell, J. L. Lebowitz, S. Baer, and W. Theumann, J. Math.
Phys. {\bf 7}, 1532 (1966).
\end{references}
\end{document}